\documentclass[twocolumn,dvipsnames]{aastex62}
\usepackage{cleveref}
\usepackage{xcolor}

\graphicspath{{./}{figures/}}

\submitjournal{AJ}

\shorttitle{\textit{Spitzer} Phase Curves of Qatar-1b}
\shortauthors{Keating et al.}

\begin{document}

\title{Smaller than expected bright-spot offsets in \textit{Spitzer} phase curves of the hot Jupiter Qatar-1b}

\correspondingauthor{Dylan Keating}
\email{dylan.keating@mail.mcgill.ca}

\author{Dylan Keating}
\affiliation{Department of Physics, McGill University, Montr\'{e}al, QC H3A 2T8, Canada}

\author{Kevin B. Stevenson}
\affiliation{John Hopkins APL, 11100 Johns Hopkins Rd, Laurel, MD 20723, USA}
\affiliation{Space Telescope Science Institute, 3700 San Martin Dr, Baltimore, MD 21218, USA}

\author{Nicolas B. Cowan}
\affiliation{Department of Physics, McGill University, Montr\'{e}al, QC H3A 2T8, Canada}
\affiliation{Department of Earth \& Planetary Sciences, McGill University, Montr\'{e}al, QC H3A 2T8, Canada}

\author{Emily Rauscher}
\affiliation{Department of Astronomy, University of Michigan, 1085 S. University Avenue, Ann Arbor, MI 48109, USA}

\author{Jacob L. Bean}
\affiliation{Department of Astronomy \& Astrophysics, University of Chicago, Chicago, IL 60637, USA}

\author{Taylor Bell}
\affiliation{Department of Physics, McGill University, Montr\'{e}al, QC H3A 2T8, Canada}

\author{Lisa Dang}
\affiliation{Department of Physics, McGill University, Montr\'{e}al, QC H3A 2T8, Canada}

\author{Drake Deming}
\affiliation{Department of Astronomy, University of Maryland, College Park, MD 20742, USA}

\author{Jean-Michel D\'{e}sert}
\affiliation{Anton Pannekoek Institute for Astronomy, University of Amsterdam, 1090 GE Amsterdam, Netherlands}

\author{Y. Katherina Feng}
\affiliation{Department of Astronomy and Astrophysics, University of California, Santa Cruz, CA 95064, USA}

\author{Jonathan J. Fortney}
\affiliation{Department of Astronomy and Astrophysics, University of California, Santa Cruz, CA 95064, USA}

\author{Tiffany Kataria}
\affiliation{Jet Propulsion Laboratory, California Institute of Technology, 4800 Oak Grove Drive, Pasadena, CA, USA}

\author{Eliza M.-R. Kempton}
\affiliation{Department of Astronomy, University of Maryland, College Park, MD 20742, USA}

\author{Nikole Lewis}
\affiliation{Department of Astronomy and Carl Sagan Institute, Cornell University, 122 Sciences Drive, Ithaca, NY 14853, USA}

\author{Michael R. Line}
\affiliation{School of Earth and Space Exploration, Arizona State University, Tempe, AZ 85281, USA}

\author{Megan Mansfield}
\affiliation{Department of Geophysical Sciences, University of Chicago, Chicago, IL 60637, USA}

\author{Erin May}
\affiliation{John Hopkins APL, 11100 Johns Hopkins Rd, Laurel, MD 20723, USA}

\author{Caroline Morley}
\affiliation{Department of Astronomy, University of Texas at Austin, Austin, TX 78712, USA}

\author{Adam P. Showman}
\affiliation{Sadly passed away while this manuscript was under review. Rest in peace, Adam.}

\begin{abstract}

We present \textit{Spitzer} full-orbit thermal phase curves of the hot Jupiter Qatar-1b, a planet with the same equilibrium temperature---and intermediate surface gravity and orbital period---as the well-studied planets HD 209458b and WASP-43b. We measure secondary eclipse of $0.21 \pm 0.02 \%$ at $3.6~\mu$m and $0.30 \pm 0.02 \%$ at $4.5~\mu$m, corresponding to dayside brightness temperatures of $1542^{+32}_{-31}$~K and $1557^{+35}_{-36}$~K, respectively, consistent with a vertically isothermal dayside. The respective nightside brightness temperatures are $1117^{+76}_{-71}$~K and $1167^{+69}_{-74}$~K, in line with a trend that hot Jupiters all have similar nightside temperatures. We infer a Bond albedo of $0.12_{-0.16}^{+0.14}$
and a moderate day-night heat recirculation efficiency, similar to HD 209458b. General circulation models for HD 209458b and WASP-43b predict that their bright-spots should be shifted east of the substellar point by tens of degrees, and these predictions were previously confirmed with \textit{Spitzer} full-orbit phase curve observations. The phase curves of Qatar-1b are likewise expected to exhibit eastward offsets. Instead, the observed phase curves are consistent with no offset: $11^{\circ}\pm 7^{\circ}$ at $3.6~\mu$m and $-4^{\circ}\pm 7^{\circ}$ at $4.5~\mu$m. The discrepancy in circulation patterns between these three otherwise similar planets points to the importance of secondary parameters like rotation rate and surface gravity, and the presence or absence of clouds, in determining atmospheric conditions on hot Jupiters. 
\end{abstract}


\section{Introduction} \label{sec:intro}

Qatar-1b is a short-period gas giant (hot Jupiter), discovered with the Qatar Exoplanet Survey  \citep{Alsubai2011}.
Its mass, radius, and orbital period are all intermediate between those of the well-studied hot Jupiters WASP-43b and HD 209458b. The three planets all have the same equilibrium temperature (see \Cref{table:comparison}). WASP-43b and HD 209458b were predicted and observed to have eastward phase curve bright-spot offsets, suggesting the presence of superrotating equatorial jets in their atmospheres \citep{Showman2009,Zellem2014,Stevenson2014,Stevenson2017,Mendonca2018, Morello2019,Kataria2015}. If the stellar flux a planet receives is what primarily determines its atmospheric dynamics, we expect the circulation of Qatar-1b to be similar to WASP-43b and HD 209458b.

\begin{deluxetable*}{ccccccc}

\tablecaption{Summary of planetary properties for HD 209458b \citep{Stassun2017}, Qatar-1b \citep{Collins2017}, and WASP-43b \citep{Esposito2017}. Qatar-1b's physical properties are intermediate between those of HD 209458 and WASP-43, except for the metallicity, which is somewhat greater than HD 209458, but consistent with WASP-43.   \label{table:comparison}}

\tablehead{\colhead{Planet Name} & \colhead{Equilibrium Temperature} & \colhead{Period} & \colhead{Mass} & \colhead{Radius} & \colhead{Surface Gravity} & \colhead{Stellar Metallicity} \\ 
\colhead{} & \colhead{(K)} & \colhead{(Days)} & \colhead{($M_{\rm Jup}$)} & \colhead{($R_{\rm Jup}$)} & \colhead{($m s^{-2}$)} & \colhead{[Fe/H]} }

\startdata
HD 209458b & $1412 \pm 64$ & 3.52 & $0.73\pm0.04$ & $1.39\pm 0.02$ & $9.79_{-0.59}^{+0.61}$ &$0.0 \pm 0.05$  \\
\textbf{Qatar-1b} &  $1418\pm27$ & 1.42 & $1.294_{-0.049}^{+0.052}$ & $1.143_{-0.025}^{+0.026}$ & $25.677_{-1.489}^{+1.577}$ & $0.171_{-0.094}^{+0.097}$ \\
WASP-43b & $1427 \pm 19$  & 0.83 & $1.998\pm 0.079$ & $1.006\pm{0.017}$ & $51\pm{11}$ &$0.05\pm0.170$ \\
\enddata

\end{deluxetable*}
The dayside brightness temperatures of Qatar-1b has been measured previously. Secondary eclipse measurements in the $K_s$ band implied an unusually high dayside brightness temperature of $1885^{+212}_{-168}$~K, which taken at face value suggests negligible day-night heat redistribution for this planet \citep{Cruz2016}. Secondary eclipse measurements with \textit{Spitzer}, combined with the $K_s$ band eclipse depth, yielded a dayside effective temperature of $1506\pm71$~K \citep{Garhart2018}, which allows for a modest degree of heat recirculation. However, full-orbit phase curve observations are the only way to quantify the day-night heat recirculation, due to the degeneracy between recirculation efficiency and albedo when interpreting eclipse only observations.

In this work, we present \textit{Spitzer} full-orbit phase curves for Qatar-1b, at 3.6$~\mu$m and 4.5$~\mu$m. From the phase curves we calculate the dayside and nightside temperatures, and in turn obtain an estimate of the Bond albedo and day-night heat recirculation efficiency \citep{Cowan2011a}.

\section{Observations and Data Analysis}
The observations consist of two full-orbit phase curves of Qatar-1b, taken with the IRAC instrument on board the \textit{Spitzer Space Telescope} \citep{Fazio2004}. Phase variations were observed at 3.6$~\mu$m on April 28, 2018, and at 4.5$~\mu$m on May 2, 2018 (PID 13038, PI: Kevin Stevenson). Both used 2 s exposure times. We performed two parallel analyses using two completely independent pipelines. The first analysis used the Photometry for Orbits, Eclipses, and Transits (POET) pipeline \citep{Stevenson2012b, Cubillos2013}, and the second used the Spitzer Phase Curve Analysis (SPCA) pipeline \citep{Dang2018, Bell2019}. Since the astrophysical signal is often buried 2--3 orders of magnitude below detector systematics, \textit{Spitzer} results have sometimes been debated \citep{Hansen2014,Schwartz2017}. While several of the most common IRAC detector systematics models produce accurate, repeatable eclipse depths \citep{Ingalls2016}, such a study has never been performed for full-orbit phase curves. It is therefore becoming common to use multiple pipelines when analyzing phase curve observations \citep{Dang2018,Bell2019,Mansfield2019}. We followed previous analyses as closely as possible to ensure a fair comparison between both pipelines. 

\subsection{Photometry}
\subsubsection{SPCA}
SPCA first performs $4\sigma$ outlier rejection to flag frames in a given \textit{Spitzer} datacube containing an outlier pixel within a $5 \times 5$ box centered on the pixel (15,15). Next, the pipeline performs frame-by-frame background subtraction by taking the median pixel value of each frame, excluding a $7\times 7$ pixel box centered on the pixel (15, 15). The pipeline uses aperture photometry to sum the remaining flux, and estimates the centroids using either the flux weighted mean (FWM) of each frame or by fitting a 2D Gaussian. 

\subsubsection{POET}
POET first performs frame by frame outlier rejection using two-iteration, four-sigma clipping to flag bad pixels. Next it fits a 2D gaussian to find the centroid of each frame, and uses the centroids to perform $5\times$ interpolated aperture photometry. The background flux is calculated using an annulus with an inner ring of 7 pixels and an outer ring of 15 pixels, and subtracted from the total.
\subsection{Centroiding}
The biggest source of noise in \textit{Spitzer} observations of bright, transiting planets is detector systematics. For IRAC Channels 1 and 2 ($3.6~\mu$m and $4.5~\mu$m) this is due to intrapixel sensitivity variations coupled with changes in target centroids. The IRAC detector pixels are not uniformly sensitive, and during each observation, the target drifts slightly across them. By now this effect is well studied, and several of the most commonly used systematics models were shown to produce repeatable, accurate eclipse depths for both real and synthetic eclipse data \citep{Ingalls2016}. Most schemes use the flux centroids to model the intrapixel sensitivity fluctuations and subtract them from the signal, which makes it crucial to obtain the correct centroids. Flux weighted mean and Gaussian centroiding are the two most commonly used methods.

In tests with synthetic data of the IRAC $3.6 \mu$m and $8 \mu$m detectors, Gaussian centroiding was shown to be more precise than the flux weighted mean method \citep{Lust2014}. However, this does not necessarily apply to every data set. In particular, Gaussian centroiding performs poorly on asymmetric point response functions. Additionally, fitting a Gaussian can introduce noise from the fitting process, whereas the flux weighted mean is computed arithmetically. 

\begin{figure}
\plotone{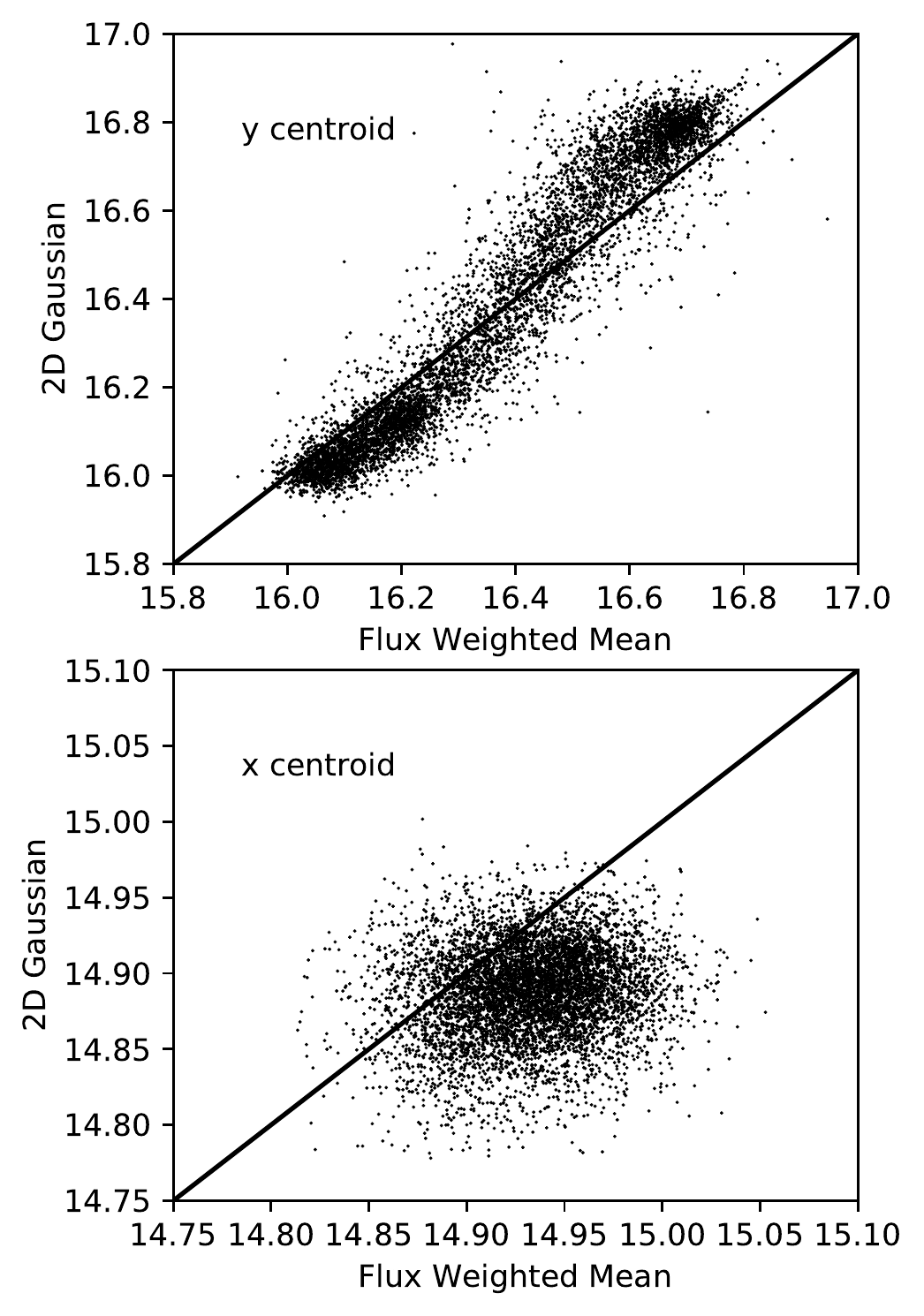}
\caption{Centroid locations, in pixels, for the $3.6~\mu$m observations, computed using two different algorithms. (The behaviour is qualitatively similar for $4.5~\mu$m observations.) The black line is what we would expect if both algorithms gave the same centroid values. Most of the telescope drift is in the $y$ direction. Although both centroiding schemes give similar raw flux, the centroid locations differ significantly, indicating that the PSF shape and asymmetry change over the course of the observation. 
\label{fig:centroids}}
\end{figure}

In our SPCA analysis, we tested both 2D Gaussian centroiding and flux weighted mean centroiding, using a range of apertures in increments of 0.25 pixels, as well as both fixed and moving apertures. To choose the aperture size for each centroiding method, we calculated the RMS scatter between the raw photometry, and a boxcar smoothed version of the photometry with a width of 10 datacubes ($\sim$ 21 minutes). We selected the aperture size with the lowest RMS scatter. In all cases, we found that apertures centered on the derived centroids (moving apertures) produced the least RMS scatter. For the $4.5~\mu$m observations, a 3.25 pixel radius aperture centered using flux weighted mean, gave the least RMS scatter. For the $3.6~\mu$m observations, a 4.25 pixel radius aperture using 2D Gaussian centroiding gave the least RMS scatter. For a given channel and aperture size, the flux difference between centroiding schemes was less than the scatter in the raw flux, meaning both centroiding schemes gave essentially the same raw flux. However, both schemes yielded  significantly different centroid locations (see \cref{fig:centroids}).

Most POET analyses have used Gaussian centroiding, with $5\times$ interpolated aperture photometry  \citep{Stevenson2017,Kreidberg2018,Mansfield2019}. The metric to pick the aperture size is slightly different than SPCA: with POET, we selected the aperture size that minimized the standard deviation of the normalized residuals (SDNR) of the phase curve fit. For the $4.5~\mu$m observations this was a 2.0 pixel radius moving aperture, and for $3.6~\mu$m it was a 2.25 pixel radius moving aperture. The reason for the different preferred aperture sizes between SPCA and POET is because POET used interpolated photometry, and because the metric to select the aperture size is different between the two pipelines. Regardless, we get qualitatively similar raw flux using both pipelines. The median background levels are 17\% and 4\%, respectively, of the median flux uncertainty at Channels 1 and 2 for our preferred aperture sizes.

\section{Modeling the Phase Curve}
The observed flux variations of the system consist of two parts: the astrophysical signal of interest, and detector systematics. We modeled and fit for these simultaneously, as \begin{equation}
F_{\mathrm{model}}(t)=A(t) \times \tilde{D}(t),
\end{equation} where $A(t)$ represents the astrophysical signal, and $\tilde{D}(t)$ is the normalized detector model.

The astrophysical signal itself has the form \begin{equation}
A(t)=F_{*}(t)+F_{p}(t),
\end{equation} where $F_{*}(t)$ is the stellar flux, and $F_{p}(t)$ is the planetary signal. With SPCA we used BATMAN \citep{Kreidberg2015} to model the occultations ---transits and secondary eclipses--- assuming a quadratic limb darkening law. The apparent stellar brightness is 
\begin{equation}
    F_{\star}(t) = T(t),
\end{equation}
where $T(t)$ is the transit light curve. Outside of transit, we assume that the stellar flux is constant.

The planetary flux ratio is \begin{equation}
F_{p}(t)=F_{\mathrm{day}} \Phi(\psi(t)),
\end{equation} where $F_{\mathrm{day}}$ is the secondary eclipse depth, $\Phi(\psi(t))$ is the phase variation of the planet, and $\psi(t)$ is the orbital phase, given by 
$\psi(t)=2 \pi\left(t-t_{e}\right) / P$ where $t_{e}$ is the time of secondary eclipse, and $P$ is the orbital period of the planet. Reflected light is negligible in the IRAC bandpasses, so we assume any light coming from the planet is thermal emission. We modelled the thermal phase variations as
\begin{equation}
\Phi(\psi)=1+C_{1}(\cos (\psi)-1)+C_{2} \sin (\psi),
\end{equation}
and imposed a prior that both the phase curve and the implied brightness map must be non-negative \citep{Cowan2008, Keating2017}. We kept $P$ fixed to 1.4200242 days \citep{Collins2017}.

With POET, we modelled the phase variations in an equivalent way: 
\begin{equation}
    \Phi(t)=1+C_{1}\cos\left[\frac{2\pi(t-t_{e})}{P}\right].
\end{equation}

We tested higher order sinusoids, but found that including them led to overfitting. We also tested fits with and without an additional linear trend in time, which can account for additional instrumental systematics or stellar variability.

\subsection{Detector Systematics}
The detector systematics, $\tilde{D}(t),$ are primarily due to intrapixel sensitivity variations. With SPCA, we considered two methods that use the flux centroid locations to model the sensitivity of the detector. The first was to fit an $n$-th degree 2D polynomial, as a function of the $x$ and $y$ position of the centroids. We tested polynomials with order 2 through 7, and used the Bayesian Information Criterion (BIC) to select the best model---the one that fits the data best without overfitting \citep{Schwarz1978}. 

The second detector model we used was BiLinear Interpolated Subpixel Sensitivity (BLISS) mapping, a non-parametric detector model. BLISS has been used successfully to analyze many phase curves \citep{Stevenson2014,Stevenson2017, Kreidberg2018,Beatty2019,Bell2019,Mansfield2019} and performed well in the \textit{Spitzer} data challenge \citep{Ingalls2016}.

We fit several combinations of detector models and astrophysical signals to the observations in each channel. First, we used Levenberg-Marquardt optimization to find the best-fit parameters, and then sampled parameter space using a Markov-Chain Monte-Carlo to obtain error estimates. We computed the BIC, and used this to select the preferred astrophysical model.

Because BLISS calculates the detector sensitivity directly, rather than letting sensitivity vary as a jump parameter, it cannot be directly compared to the polynomial models using the BIC. However, the BIC can be used to select the preferred astrophysical model for a given BLISS implementation.

We also tested pixel-level decorrelation (PLD), which does not explicitly assume a functional form for the detector sensitivity and does not use centroids \citep{Deming2015}. Although first-order PLD is inadequate when the stellar centroid moves more than about 0.1 pixel, second-order PLD has been used successfully for phase curve observations \citep{Zhang2018,Bell2019}. In both of our observational channels the centroids move by nearly a pixel over the course of the observation, so we implemented second-order PLD. As PLD is parametric, it can be directly compared to the polynomial model using BIC.

For the POET analyses, we used a BLISS detector model. Because BLISS is non-parametric, it is flexible and has the advantage of running quickly in a Monte Carlo.

Lastly, because of the differences in photometry and sigma-clipping between pipelines, the respective datasets are not exactly the same between the two analyses. There is no perfect way to compare analyses between two different pipelines. One way is to compare which one of them gives the lowest fit residuals, and another is to compare the log-likelihoods per datum of the models \citep{Bell2019}.

\begin{table*}[t]
\centering
\caption{Summary of key light curve parameters for both wavelengths. Our fiducial analysis is the POET fit to the unbinned data using a BLISS detector model, as it produced the smallest fit residuals.  \label{table:summary}}
\begin{tabular}{ccccccc}

Wavelength & Eclipse Depth ($\%$) & $R_{\rm p}/R_{\star}$& Amplitude (ppm) & Phase Offset & $T_{\rm bright, day}$ (K) & $T_{\rm bright, night}$ (K)  \\
 \hline \\ [-2.5ex]
$3.6~\mu$m &$0.21 \pm 0.01 $  & $0.144\pm0.001$&$660\pm91$ &  $11^{\circ} \pm 7^{\circ}$ & $1542_{-31}^{+32}$  &$1167_{-74}^{+69}$ \\
$4.5~\mu$m &  $0.30 \pm 0.02 $ & $0.145\pm0.001$ & $918\pm114$ & $-4^{\circ} \pm 7^{\circ}$ & $1557_{-36}^{+35}$ &$1117_{-71}^{+76}$ \\
 \hline \\
 [-1.5ex]
\end{tabular}
\end{table*}

\subsection{Binning}
Some phase curve analyses have fitted binned data while others have fitted the unbinned data. There are arguments for and against binning. Binning data before fitting filters out high frequency noise, improves centroid position accuracy, and makes the fits run faster, among other advantages \citep{Deming2015}. However, it can also distort the light curve shape if the bin size is too large \citep{Kipping2010}. Binning can also smooth over short timescale telescope pointing variations. To date, there has been no systematic study of the effects of bin size on retrieved phase curve shapes. There is some preliminary evidence, however, that coarse binning yields phase curve shapes discrepant with results from fitting unbinned data. This effect will be fully explored in upcoming work (May et al. 2020, in prep.)

Fitting unbinned data with SPCA was prohibitively slow, especially when testing multiple model combinations and higher order polynomial and PLD models. Therefore we binned the observations by datacube (64 frames, or 128 s) which is much shorter than the occultations, and equal to the bin sizes used in previous work with SPCA \citep{Dang2018,Bell2019}. 

With POET we were able to fit the entire unbinned dataset, because POET is optimized to run with multiprocessing. Best-fit parameters and uncertainties from the two pipelines are shown in \Cref{table:ch2centroids,table:ch1centroids}. 

\section{Results}
The intrapixel sensitivity variations are less severe for the $4.5~\mu$m channel and typically easier to fit than the $3.6~\mu$m channel, so we begin by summarizing the $4.5~\mu$m results. See \Cref{table:summary} for a summary of the fiducial light curve parameters at both wavelengths.

\subsection{$4.5~\mu$m Observations}
\begin{figure*}[t]
\plotone{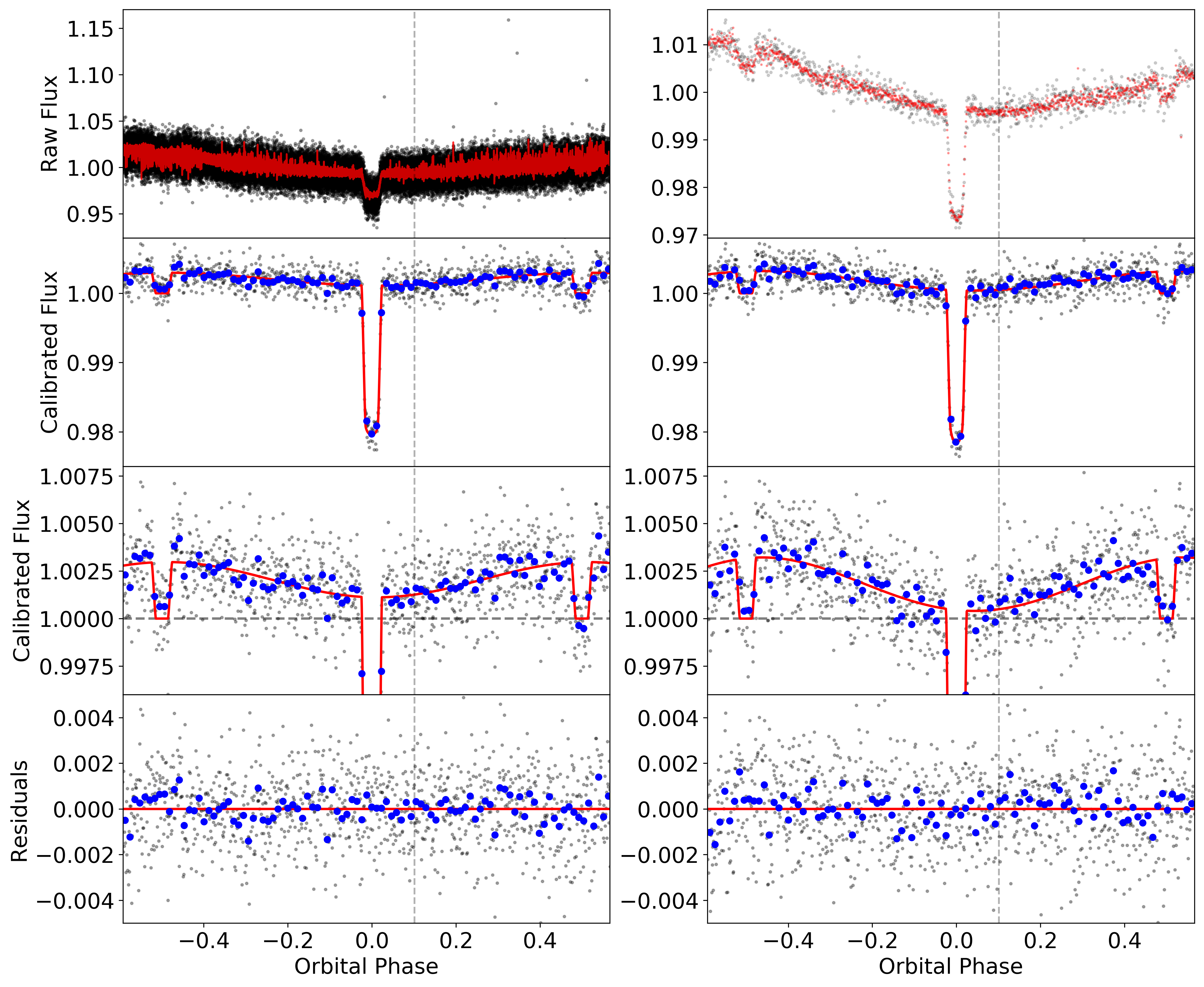}
\caption{\textit{Left:} Raw photometry and light curve model for the $4.5~\mu$m observations of Qatar-1b using the POET pipeline. We fitted the unbinned dataset, shown in black in the top panel, but bin the data by datacube when plotting below for clarity. \textit{Right:} Results from the SPCA pipeline. The best fit combined astrophysics $\times$ detector model for each pipeline is shown in red in the top panels. The grey dots come from binning the data by datacube (64 frames), and the blue dots are more coarsely binned.  The red line is the final best fit model for each. The two middle panels show the best-fit astrophysics model, with the detector systematics removed. The bottom panels show the residuals of the best-fit light curve and detector systematics models subtracted from the raw signal. The dashed line indicates where the Astronomical Observing Request break occurs.
\label{ch2Both}}
\end{figure*}

To see if using different centroid algorithms affected the fitted parameters, we fit the photometry obtained from both algorithms with SPCA. The flux weighted mean (FWM) centroiding gave lower scatter in the residuals, as well a higher log-likelihood than Gaussian centroiding. A first order sinusoid, and no linear slope, was the preferred astrophysical model. All combinations of detector models and centroiding algorithms produced consistent eclipse depths within the error bars. In all the fits we tried, we found a slight westward phase offset. The phase curve using the second order polynomial detector model is shown in the righthand column of \Cref{ch2Both}. 

Our POET analysis used a BLISS detector model. We selected the knot spacing such that bilinear interpolation performed better than nearest neighbour interpolation \citep{Stevenson2012a}, which was 0.019 pixels in each direction. Again we found that a first order sinusoid, with no linear slope, gave the lowest BIC and lowest residuals in the final fit compared to other astrophysical models. We attempted decorrelating against the PSF width, but it gave a higher BIC for this wavelength.

\begin{figure}[t]
\plotone{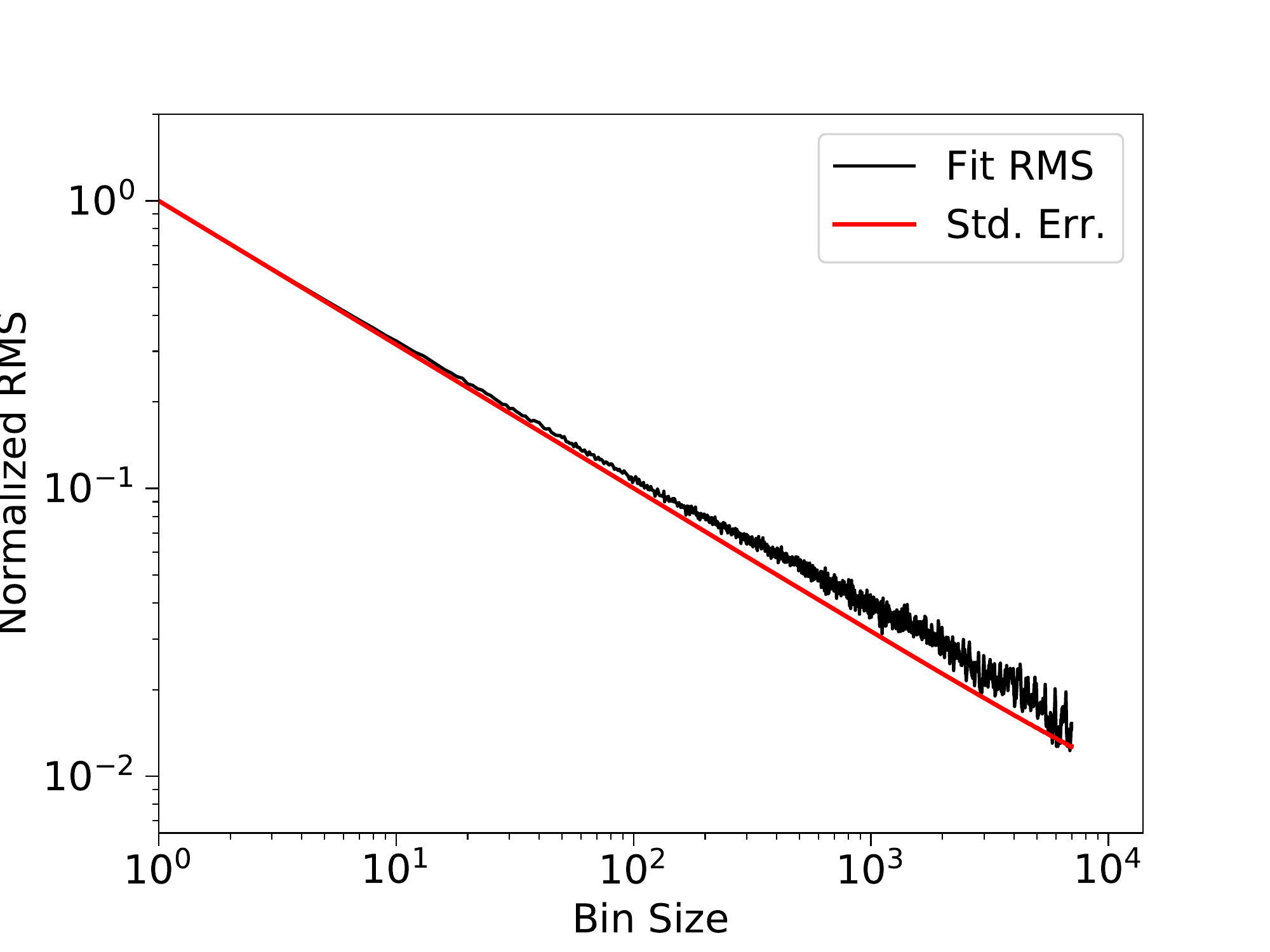}
\caption{Root-mean-squared residuals versus bin size for the $4.5~\mu$m phase curve fit with POET. The red line is the expected behaviour assuming white noise.
\label{fig:ch2RedNoise}}
\end{figure}

The POET analysis yielded an eclipse depth consistent with the SPCA analysis, as well as a slightly westward phase offset of $-4^{\circ} \pm 7^{\circ}$, and lower residuals in the final fit (see \Cref{ch2Both}). The residuals were 1.11 greater than the photon noise limit. There was little red noise in the final fit (\Cref{fig:ch2RedNoise}). Using the phase curves fluxes, we get a dayside brightness temperature of $1557_{-36}^{+35}$~K, and a nightside brightness temperature of $1117_{-71}^{+76}$~K.

\begin{figure*}[t]
\plotone{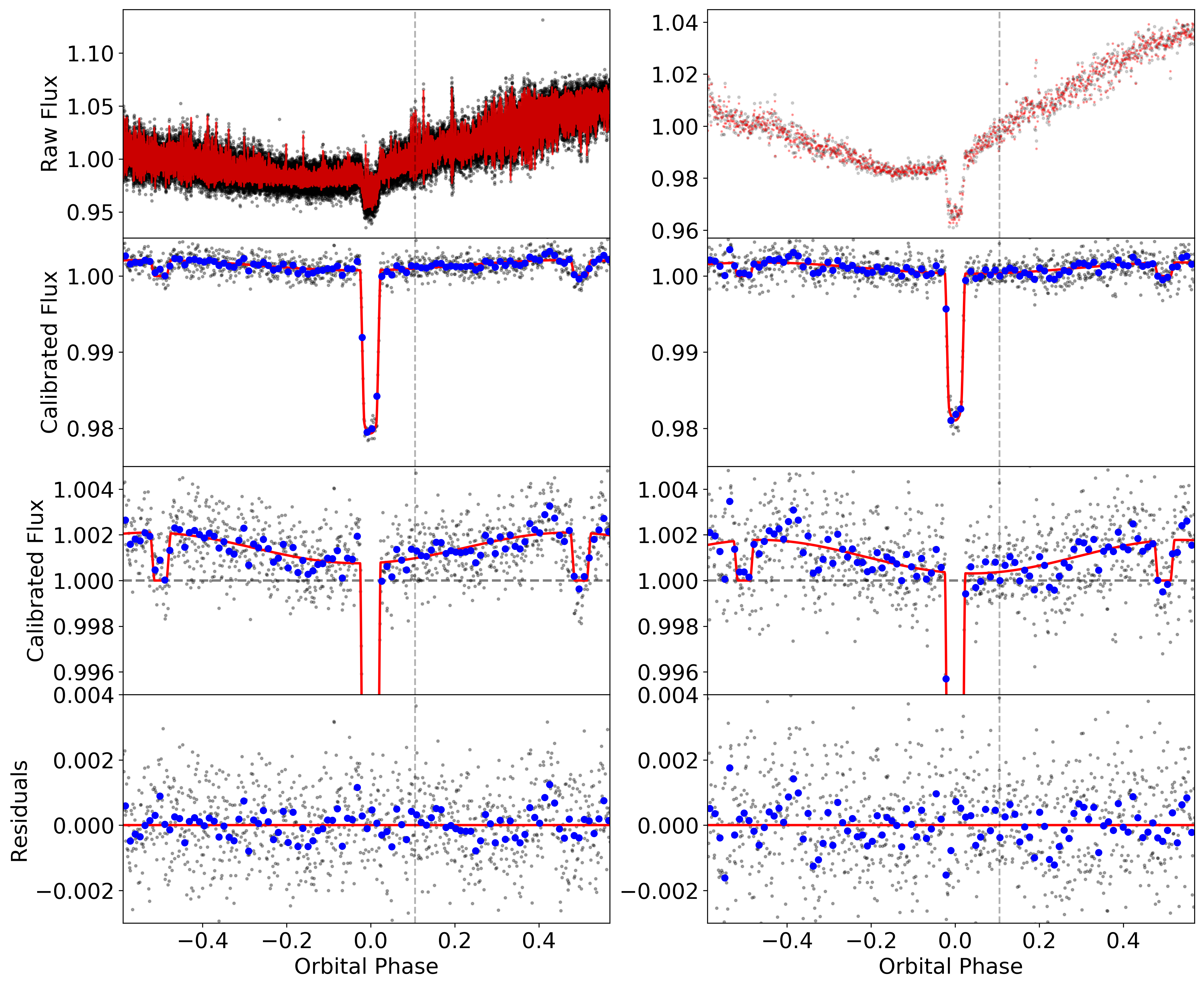}
\caption{The same figure as \Cref{ch2Both} but for the $3.6~\mu$m observations. 
\label{ch1Both}}
\end{figure*}

\subsection{$3.6~\mu$m Observations}
For the $3.6~\mu$m phase curve observations, we also tried fitting photometry using both centroiding algorithms. We again experimented with different combinations of detector polynomial orders and astrophysical models. For this channel only, decorrelating against the PSF width resulted in a dramatically lower BIC for all the polynomial and BLISS fits we tried, with both SPCA and POET. For the flux weighted mean photometry, a second order polynomial gave a lower BIC than higher orders. For the 2D Gaussian photometry, we tried polynomial orders from two to seven, and found that a sixth order polynomial gave the lowest BIC among the polynomial models. 

Unlike for the $4.5~\mu$m observations, the eclipse depths did not agree for the reductions using different centroiding algorithms. The eclipse depths from the flux weighted mean photometry were twice as deep as those using Gaussian centroiding or the one reported by \citet{Garhart2018}. The likelihoods were lower, and the residuals were higher, than the fits to the Gaussian centroiding photometry. Gaussian centroiding also gave lower scatter in the raw photometry and centroids. For these reasons we chose the Gaussian centroiding photometry as our fiducial dataset for the SPCA fits. 

We also saw a discrepancy between the different detector models: the polynomial and BLISS models gave consistent eclipse depths to one another but the phase offsets are $2.5 \sigma$ discrepant. With BLISS, an eastward phase offset was favoured, but with the polynomial detector model, a westward offset was favoured. BLISS gave a higher likelihood and lower residuals than the polynomial fit.

Because of the discrepancy in the inferred parameters, we tried a centroid-agnostic detector model: a second-order PLD using a $5\times5$ grid of pixels. We performed the decorrelation using the individual pixel fluxes, but used the 2D Gaussian aperture photometry as our dataset, rather than the sum of pixels, in keeping with past analyses \citep{Deming2015, Zhang2018, Bell2019}. PLD performed better than the polynomial models, achieving a higher log likelihood, lower BIC, and lower residuals in the final fit. The fitted phase curve is shown in the righthand column of \Cref{ch1Both}. The PLD fit gave a phase curve offset consistent with that of the sixth order polynomial fit. 

\begin{figure}
\plotone{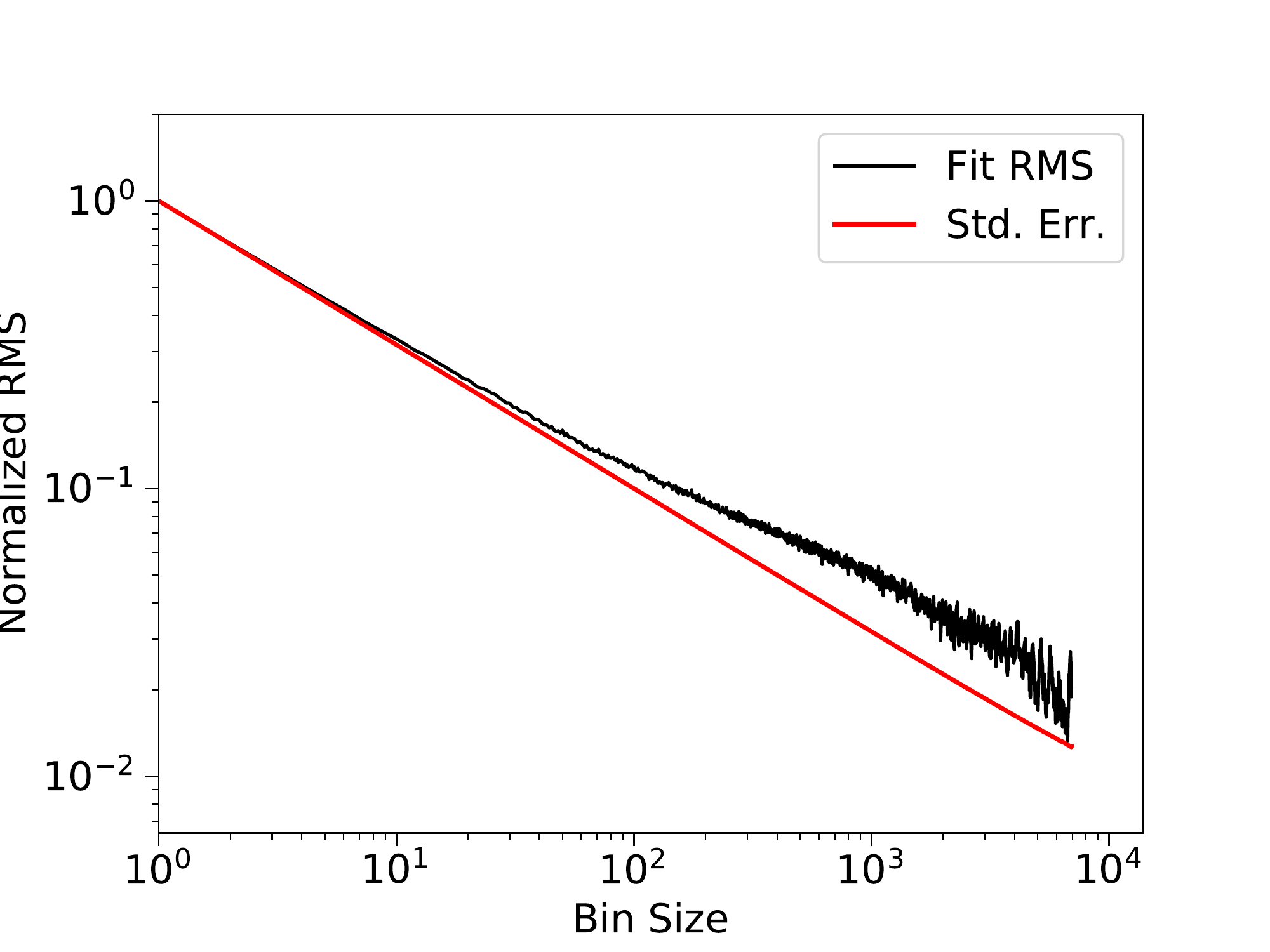}
\caption{The same plot as \cref{fig:ch2RedNoise} but for the $3.6~\mu$m phase curve.
\label{fig:ch1RedNoise}}
\end{figure}

Our POET analysis again used a BLISS detector model, with an ideal knot spacing of 0.012. A first order sinusoid, with a linear ramp, and a fit to the PSF width was the preferred model. The fit is shown in the left panel of \Cref{ch1Both}. We found an eastward offset of $11^{\circ} \pm 7^{\circ}$. This is $2.3\sigma$ away from the offset found with SPCA. Because the POET analysis used the full, unbinned dataset, and yielded lower residuals than SPCA, we take that as our fiducial analysis for the rest of this paper. The residuals were 1.16 times greater than the photon noise limit. The final fit removed most of the red noise (\Cref{fig:ch1RedNoise}). The dayside and nightside brightness temperatures are $1542_{-31}^{+32}$~K, and $1167_{-74}^{+69}$~K.

\begin{table*}
\centering
\caption{Comparison of the best $4.5~\mu$m phase curve fit for each pipeline. SDNR stands for standard deviation of the normalized residuals, for which we show the residuals binned by datacube (128~s) for both fits. The fit using the POET pipeline gave the lowest residuals. The transit and eclipse depths are consistent within $1\sigma$ between both analyses, but the phase offsets and amplitudes are not. \label{table:ch2centroids}}
\begin{tabular}{ccccccccc}

Pipeline & Centroids & Bin Size & Detector                &Eclipse Depth ($\%$) & $R_{\rm p}/R_{\star}$ & Amplitude (ppm)& Offset & SDNR (ppm)\\
 \hline \\ [-1.5ex]
POET &Gaussian    &     2s      & BLISS          &  $0.30\pm0.02$ &$0.1453(8)$& $918\pm114$  &$-4^{\circ} \pm 7^{\circ}$ & 1140 \\
SPCA & FWM    &  128s &Poly 2     &   $0.31\pm0.02$ &$0.146(1)$& $1336\pm 101$ &$-17^{\circ} \pm 4^{\circ}$ & 1660 \\
 \hline \\
 [-1.5ex]
\end{tabular}
\end{table*}




\begin{table*}
\centering
\caption{The same as \cref{table:ch2centroids} but for $3.6~\mu$m \label{table:ch1centroids} The transit depths and phase amplitudes are consistent within $1\sigma$ between both analyses, but the transit depths and phase offsets are not.}
\begin{tabular}{ccccccccc}

Pipeline & Centroids & Bin Size &Detector &  Eclipse Depth ($\%$) & $R_{\rm p}/R_{\star}$ & Amplitude (ppm) & Offset & SDNR (ppm) \\
 \hline \\ [-1.5ex]
POET &Gaussian  & 2s             & BLISS &$0.21\pm 0.01 $ & $0.1443(6)$ &$660\pm91$ & $11^{\circ} \pm 7^{\circ}$& 857  \\
SPCA & Gaussian  & 128s            & PLD 2   &$0.18\pm 0.02 $ &  $0.137(1)$ &$675\pm 134$ & $-22^{\circ} \pm 18^{\circ}$ & 1448 \\

 \hline \\ [-1.5ex]
\end{tabular}
\end{table*}

\section{Discussion}
The two main observational quantities calculated from thermal emission phase curves are the amplitude of variations, and the phase at which the peak flux occurs. These are, respectively, measures of the temperature differences between the day and night hemispheres of the planet, and the ability of the planetary winds to advect hot gas away from the substellar point before it can cool. There is a large body of literature examining expectations for hot Jupiter atmospheric circulation patterns and how these physics translate into observed thermal phase curves (e.g. see reviews by \citet{Parmentier2018} and \citet{Heng2015}). In a comprehensive study, \cite{Komacek2017} show that to first order, the day-night contrast on a planet should increase with increasing equilibrium temperature and the hottest region of the planet should be closer to the substellar point (i.e., smaller phase offsets). Higher order effects include the planet's rotation rate and its gravity. Moreover, if the rotation rate is too slow, it can disrupt the circulation pattern, breaking the predicted trends \citep{RauscherKempton2014}.

Any sources of drag in an atmosphere can slow down the winds, leading to smaller bright-spot offsets \citep{Komacek2017}. One obvious culprit for hot Jupiters is magnetic drag. If a planet is hot enough for thermal ionization and has an appreciable magnetic field, then ions in the atmosphere can interact with the magnetic field, acting as a source of drag. All else being equal, the effective drag strength should increase with increasing stellar irradiation \citep{Perna2010}, but it depends on magnetic field strength and metallicity. Magnetic drag could directly impact atmospheric circulation.

Clouds can also directly affect the atmospheric circulation through feedback with the temperature and wind structures themselves. They can also influence the phase curves by disconnecting the emitted flux from the temperature structure \citep{Roman2019}. When clouds are thick enough to provide significant scattering, it becomes important to model their effects within a GCM simulation rather than post-processing. If the clouds are so thin that feedback is not important,  \citet{Parmentier2016} predict that clouds should lead to observable trends in \textit{Kepler} phase curve shapes as a function of equilibrium temperature.

By comparing thermal phase curves of planets with similar equilibrium temperatures, such as Qatar-1b, HD 209458b, and WASP-43b, any differences can be attributed to differences in circulation efficiency, due to differences in rotation, gravity, atmospheric drag, or clouds. 

\subsection{Temperatures and Energy Budget}
First we consider the energy budget of Qatar-1b and compare it to HD 209458b and WASP-43b. By combining brightness temperatures at several wavelengths, it is possible to estimate the total bolometric flux emitted by a given hemisphere of a planet; this can be quantified by an effective temperature of the hemisphere. We estimated the dayside and nightside effective temperatures of Qatar-1b using Gaussian process regression, which was shown to give more robust temperature estimates than using the error weighted mean or linear interpolation \citep{Pass2019}.

The dayside effective temperature of Qatar-1b is $1588\pm73$~K, and the nightside effective temperature is $1163\pm79$~K. These include the systematic error introduced when converting from brightness temperatures to an effective temperature \citep{Pass2019}. Using these estimates, we obtain a Bond albedo of $0.12_{-0.16}^{+0.14}$ and a day-to-night heat recirculation efficiency of $0.52_{-0.11}^{+0.12}$ \citep{Cowan2011a}, confirming that Qatar-1b does in fact circulate heat from day to night. The heat recirculation efficiency is consistent with the value of $0.51_{-0.13}^{+0.15}$ for HD 209458b \citep{Keating2019}. The heat recirculation efficiency of WASP-43b is debated. It was initially reported to be negligible: $0.002_{-0.002}^{+0.01}$ \citep{Stevenson2017}, but demanding that the brightness map of WASP-43b be strictly positive gave a much higher value of $0.51\pm{0.08}$ \citep{Keating2017}. Using the reanalyzed, non-negative phase curves of \citet{Mendonca2018} gives a heat recirculation efficiency of $0.27_{-0.11}^{+0.12}$ \citep{Keating2019}. 

\begin{figure*}
\plotone{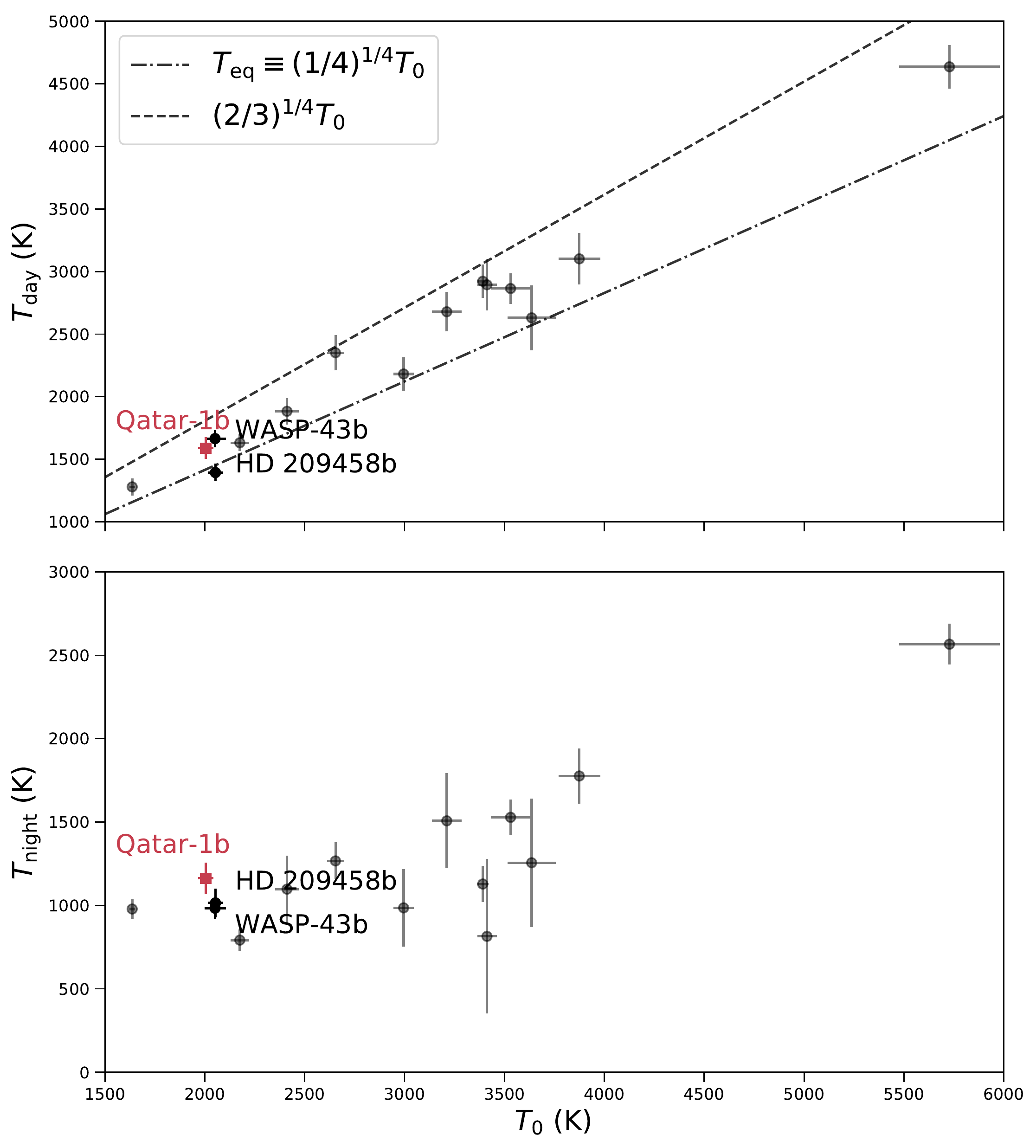}
\caption{An updated version of the dayside (top panel) and nightside (bottom panel) temperatures plot from \citet{Keating2019}, including Qatar-1b and KELT-9b \citep{Mansfield2019}. Qatar-1b, HD 209458b, and WASP-43b are shown as opaque points. The horizontal axis is the irradiation temperature, $T_{0} \equiv T_{\star} \sqrt{R_{\star} / a}$. On the top panel we also plot the equilibrium temperature $T_{\rm eq} \equiv (1/4)^{1/4} T_{0}$ (dashed-dotted line), and the dayside temperature in the limit of no heat transport: $(2/3)^{1/4} T_{0}$ (dashed line). Qatar-1b is plotted with a square marker, and fits the trend that hot Jupiters have nightside temperatures around 1100~K. Ultra-hot Jupiters, planets with irradiation temperatures above $\sim$3500~K, have hotter nightsides due to heat transport from hydrogen dissociation and recombination. 
\label{fig:TdayTnight}}
\end{figure*}

We plot the dayside and nightside temperatures along with those of all other hot Jupiters with full-orbit infrared phase curves in \Cref{fig:TdayTnight}. The nightside temperature of Qatar-1b is in line with the trend that hot Jupiters all have nightside temperatures of approximately 1100~K, likely due to nightside clouds \citep{Keating2019,Beatty2019}, which are predicted to be ubiquitous on hot Jupiters \citep{Parmentier2016}. Ultra-hot Jupiters, with irradiation temperatures greater than about 3500~K, have hotter nightsides due to additional heat transport from hydrogen dissociation and recombination. For instance, the ultra-hot Jupiter KELT-9b is so irradiated that nothing can condense, even on its nightside \citep{Mansfield2019}.

\subsection{Phase Offsets}

We detected no phase offset at $4.5~\mu$m ($4 \pm 7^{\circ}$, westward). This stands in stark contrast to the eastward offsets of $40 \pm 6^{\circ}$ and $21.1 \pm 1.8^{\circ}$ for HD 209458b and WASP-43b, respectively, at the same wavelength \citep{Zellem2014,Stevenson2017}. Two reanalyses of WASP-43b at $4.5~\mu$m also found eastward phase offsets, of $12 \pm 3^{\circ}$ \citep{Mendonca2018} and $11.3 \pm 2.1^{\circ}$ \citep{Morello2019}. 

\begin{figure*}
\plotone{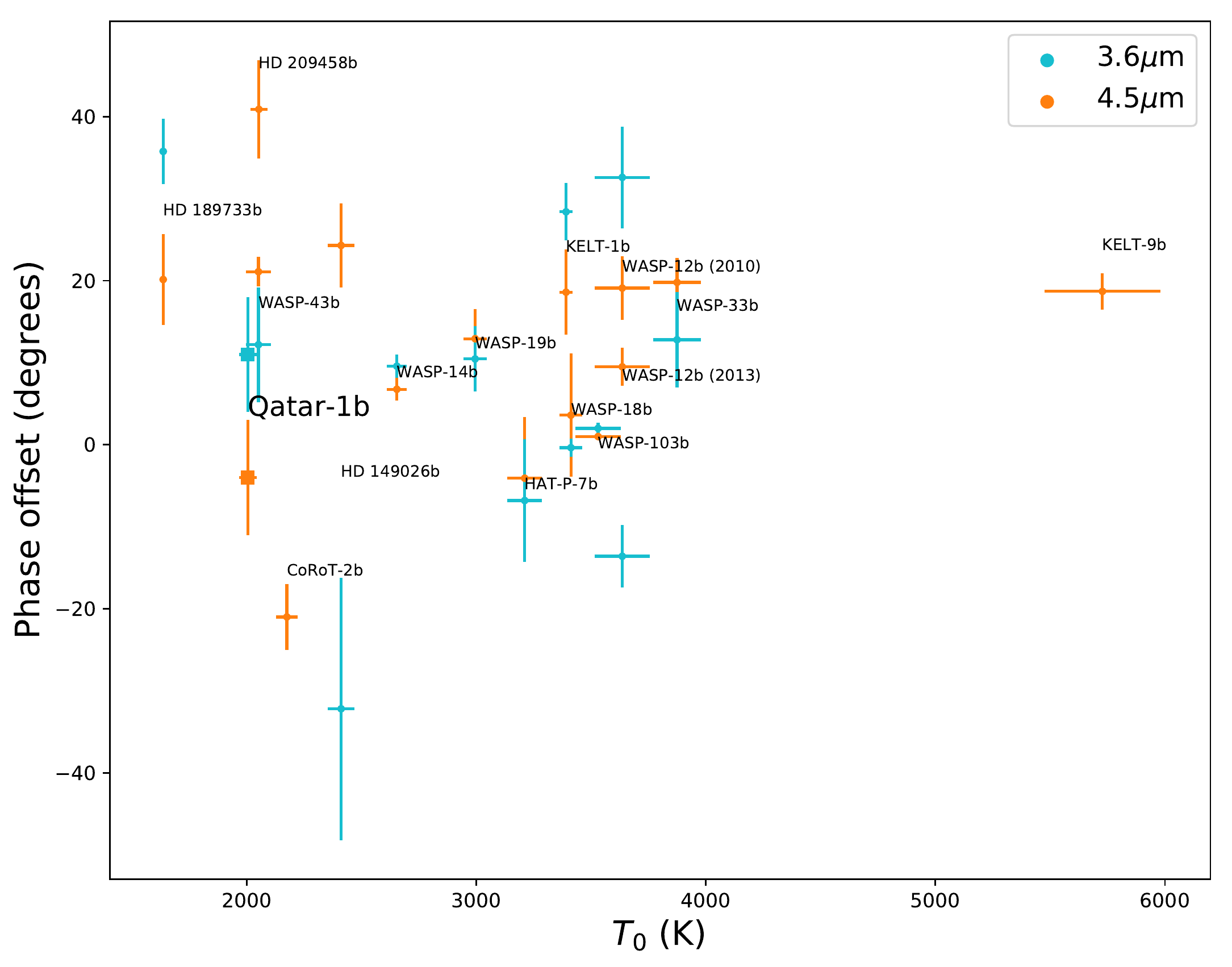}
\caption{Phase offsets for all hot Jupiters with full-orbit phase curves. Qatar-1b is denoted with a square marker.  
\label{fig:phaseoffsets}}
\end{figure*}

\begin{figure*}
\plotone{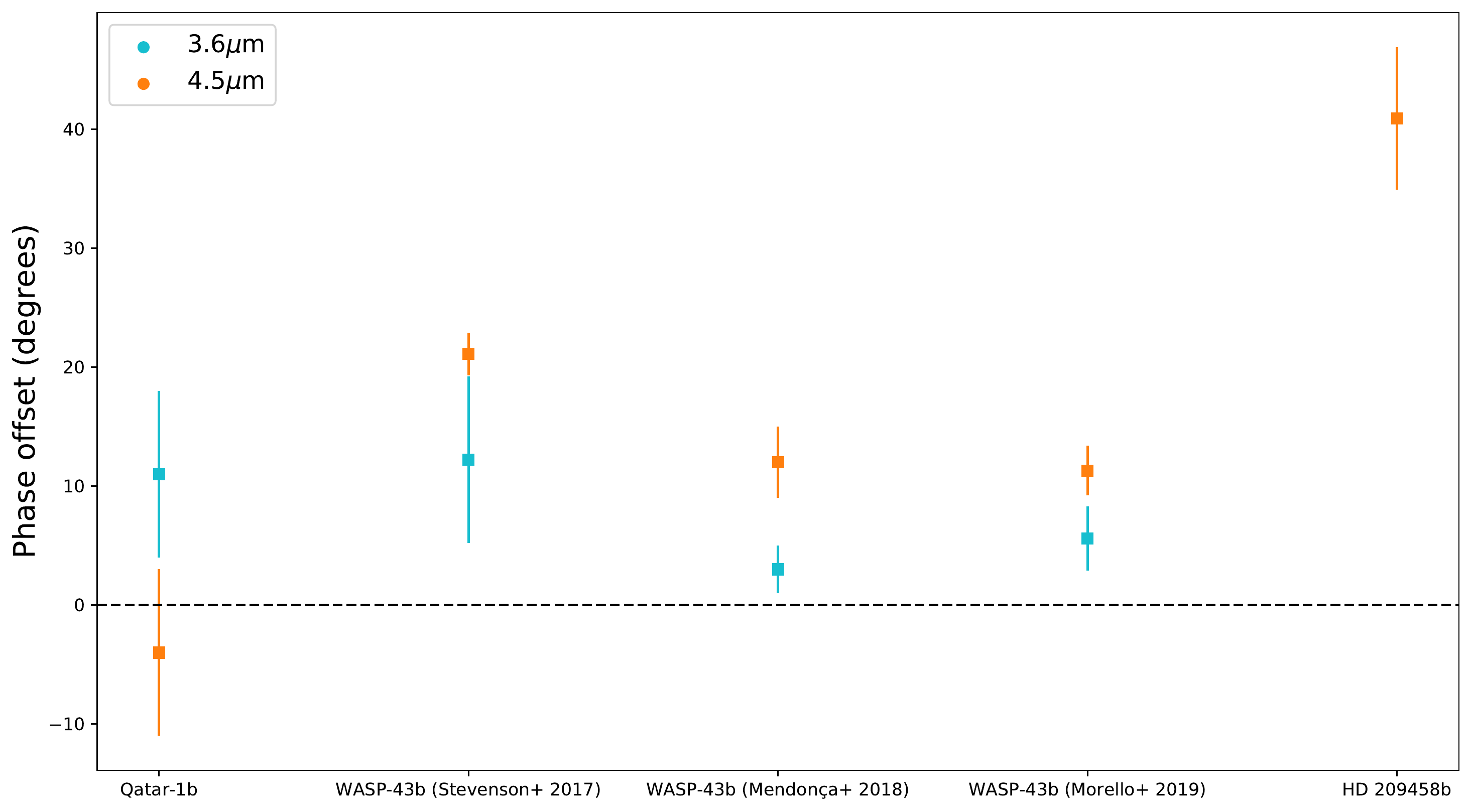}
\caption{Phase offsets for just Qatar-1b, WASP-43b, and HD 209458b using all currently published values \citep{Zellem2014,Stevenson2017,Mendonca2018,Morello2019}. 
\label{fig:threestooges}}
\end{figure*}

At $3.6~\mu$m we detected a phase offset of $11 \pm 7^{\circ}$ (eastward). This is consistent with the result of $12.2 \pm 7^{\circ}$ for WASP-43b \citep{Stevenson2017}. The reanalyzed $3.6~\mu$m offsets of WASP-43b are $3 \pm 2^{\circ}$ (eastward) \citep{Mendonca2018} and $5.6 \pm 2.7^{\circ}$ (eastward) \citep{Morello2019}. The offsets we observed for Qatar-1b in both channels are on opposite sides of the substellar point, but are just $1.5\sigma$ away from one another. We plot the phase offsets for Qatar-1b with the \textit{Spitzer} phase curve offsets for the whole suite of hot Jupiters in \Cref{fig:phaseoffsets}, and plot the phase offsets of Qatar-1b with all published phase offsets of WASP-43b and HD 20945b in \Cref{fig:threestooges}. 

The planet CoRoT-2b was the first hot Jupiter with a robustly detected westward phase curve offset \citep{Dang2018}. The offset was $21 \pm 4^{\circ} $ west at $4.5~\mu$m. A $3.6~\mu$m phase curve was not observed. The authors suggested three scenarios to cause the westward offset: non-synchronous rotation, magnetic effects, or eastern clouds.

Like other hot Jupiters, Qatar-1b is expected to be tidally locked into synchronous rotation \citep{Showman2002}. However, if it was not synchronously rotating, GCM simulations suggest Qatar-1b could have a reduced eastward offset or even westward offset \citep{Showman2009,RauscherKempton2014}. Either way, we would expect to see the same direction of phase offset in both channels. 

Magnetic drag can also reduce eastward phase offsets, or produce westward offsets \citep{Rauscher2013,Hindle2019}. At high enough equilibrium temperatures, alkali metals in a planet's atmosphere thermally ionize and interact with the planet's magnetic field, acting as a source of drag.  The ultra-hot Jupiter WASP-18b was found to have a small phase offset in its Wide Field Camera 3 phase curve, which the authors attributed to magnetic drag \citep{Arcangeli2019}. Magnetic interactions can also cause phase offsets to periodically change from east to west \citep{Rogers2017}. Such a temporal change in phase offset was observed in \emph{Kepler} phase curves of HAT-P-7b \citep{Armstrong2016}, and the $3.6~\mu$m phase curve of WASP-12b \citep{Bell2019}. 

Qatar-1b's equilibrium temperature is below the threshold where magnetic effects are expected to be significant \citep{Menou2012}, but its host star has a higher metallicity than HD~209458 and WASP-43. A higher stellar metallicity may a suggest a larger number of trace metals in the planetary atmosphere, resulting in a more strongly ionized atmosphere. Magnetic field strengths of hot Jupiters are unknown, but could potentially be orders of magnitude higher than Jupiter \citep{Yadav2017}. High metallicities and strong planetary magnetic fields can decrease the threshold for magnetic drag effects to become important \citep{Menou2012}.

The large uncertainty on the $4.5~\mu$m phase curve offset means it is also consistent with a negligible eastward offset. This may be evidence of magnetic drag. Otherwise, if the phase curve offsets at both wavelengths are truly on opposite sides of the substellar point, this could be evidence of magnetic variability on the timescale of about five days (the time between the observations). If magnetic drag is reducing the wind speed, then deep transport is needed to move heat to the nightside.

Dayside clouds could also cause reduced eastward phase offsets, westward offsets, or variable offsets. The reflected optical phase curve of the planet Kepler-7b has a westward offset, best explained by westward clouds \citep{Demory2013,Rauscher2017}. In the infrared, a westward offset could be caused by eastward clouds \citep{Dang2018}, although cloud models do not generally predict this. Optically thick dayside clouds could simply be obscuring the transported heat of Qatar-1b, leading to negligible phase offsets in both channels. The dayside brightness temperatures are the same at both wavelengths, consistent with blackbody emission from an optically thick cloud deck. Time variable cloud coverage could also cause bright-spot variations, similar to the variability seen for brown dwarfs.

The presence or lack of clouds can be tested by measuring the albedo at optical wavelengths, using the UVIS mode of the Wide Field Camera 3 instrument on board the \textit{Hubble Space Telescope}. Optical phase curves, such as one from the Transiting Exoplanet Survey Satellite (TESS), could also measure the offset at visible wavelengths. TESS has already observed the phase curve of Qatar-1b, so it could be analyzed as has been done for some hotter planets \citep{Shporer2019,WongWASP,WongKELT,Bourrier2019,Daylan2019}.

Overall, Qatar-1b makes an interesting target for the \textit{James Webb Space Telescope} (JWST), especially as a comparison to WASP-43b, which will be observed extensively by JWST as part of Early Release Science and Guaranteed Time Observations \citep{Bean2018}. The upcoming ARIEL mission will also be able to measure spectroscopic phase curves at a similar range of wavelengths as JWST \citep{Tinetti2018} and potentially probe time variability.

\section{Conclusion}
We presented full-orbit infrared phase curves of Qatar-1b taken with the \textit{Spitzer} space telescope at $3.6~\mu$m and $4.5~\mu$m. We summarize our results below. 
\begin{itemize}
  \item The dayside brightness temperatures are the same at both wavelengths. 
  \item The nightside brightness temperatures are the same at both wavelengths, and follow the trend that hot Jupiters have the same nightside temperatures \citep{Keating2019, Beatty2019}.
  \item Qatar-1b circulates a moderate amount of heat from day to night, similar to HD 209458b, but has a higher recirculation efficiency than WASP-43b. The three planets all receive the same amount of stellar irradiation.
  \item The bright-spot offsets for the two phase curves of Qatar-1b are consistent with zero. They stand in contrast to the significant eastward hotspot offsets predicted by GCMs and observed for HD 209458b and WASP-43b. The three planets all receive the same amount of stellar irradiation, so this discrepancy points to the importance of secondary parameters like rotation rate, gravity, or metallicity in determining their atmospheric conditions. Some physical mechanisms to produce the small offsets for Qatar-1b are subsynchronous rotation, magnetic effects, or dayside clouds, but there is so far no strong evidence for any of these. 
  \item Qatar-1b is an attractive target for the JWST and ARIEL missions, especially as a comparison to WASP-43b which will be observed extensively. In the meantime, \textit{Hubble} UVIS observations may be able to test the dayside cloud hypothesis, as would an analysis of the TESS optical phase curve.
\end{itemize}

\acknowledgements

This work is based on observations made with the Spitzer Space Telescope, which is operated by the Jet Propulsion Laboratory, California Institute of Technology under a contract with NASA. D.K., N.B.C., T.J.B., and L.D. are affiliated with, and supported by, the McGill Space Institute, the Institute for Research on Exoplanets, and the Centre for Research in Astrophysics of Qu\'ebec. J.M.D acknowledges the European Research Council (ERC) European Union’s Horizon 2020 research and innovation programme (grant agreement no. 679633; Exo-Atmos) and the Amsterdam Academic Alliance (AAA) Program. \software{astropy (The Astropy Collaboration 2013, 2018), matplotlib (Hunter 2007), Scipy (Jones et al. 2001), POET (Stevenson et al. 2012; Cubillos et al. 2013), SPCA (Dang et al. 2018; Bell et al. 2019)}

\end{document}